\documentclass[noshowpacs,amsmath,
twocolumn,
superscriptaddress,
8pt,
aps,prb]{revtex4-2}
\pdfoutput=1
\usepackage{setspace}
\usepackage{amsmath}
\usepackage{multirow}
\usepackage{graphicx}
\usepackage[export]{adjustbox}
\usepackage{upgreek}

\usepackage{verbatim}
\usepackage{amsfonts}
\usepackage{amssymb}
\usepackage{epstopdf}
\usepackage{dcolumn}
\usepackage{xcolor}
\usepackage{verbatim}
\usepackage[version=4]{mhchem}
\usepackage{hyperref}
\usepackage{siunitx}
\usepackage{textcomp}
\setcitestyle{super}
\DeclareGraphicsExtensions{.pdf,.eps,.png,.jpg,.mps}
\begin{document}
\title{Dispersion-managed octave soliton microcombs in heterostructured microresonators}

\author{Xinghong Li$^{1}$, Wenxin Zhang$^{1}$, Yiyang Lu$^{1}$, Zhaoyi Wang$^{1}$, Qingzhuo Xuan$^{1}$, Shangyuan Li$^{1}$, Xiaoping Zheng$^{1}$, and Xiaoxiao Xue$^{1,\dagger}$\\$^{1}$Department of Electronic Engineering, Beijing National Research Center for Information Science and Technology, Tsinghua University, Beijing 100084, China\\$^{\dagger}$Corresponding author: \href{mailto:xuexx@tsinghua.edu.cn}{xuexx@tsinghua.edu.cn}}

\begin{abstract}
Controlling group velocity dispersion is of fundamental importance in ultrafast optics, particularly for supercontinuum generation and optical frequency comb synthesis. However, the simultaneous, independent tailoring of multiple dispersion coefficients over an ultra-broad bandwidth remains a formidable challenge in conventional nanophotonic platforms. Here, we demonstrate a robust strategy for broadband dispersion management using heterostructured microresonators comprised of adiabatically concatenated waveguides with distinct geometries. By meticulously engineering the local dispersion profiles, we can flexibly synthesize the global effective dispersion coefficients of different orders, effectively expanding design degrees of freedom beyond conventional limits. As a benchmarking demonstration, we fabricate heterostructured silicon nitride microresonators using a commercial foundry process and successfully generate octave-spanning soliton microcombs with a repetition rate as low as 118 GHz. These microcombs feature deterministically tunable dispersive waves operating across the 290\textendash{}310 THz range. Such octave-spanning microcombs with detectable repetition rates and carrier-envelope offset frequencies are readily applicable to $f$-$2f$ self-referencing in optical clocks and frequency synthesizers. The proposed heterostructured architecture establishes a versatile paradigm for generating ultrawideband soliton microcombs with tailorable spectral profiles.
\end{abstract}

\maketitle

\medskip

Microcombs leverage third-order ($\chi^{(3)}$) Kerr nonlinearities to generate thousands of coherent comb lines from a single continuous-wave (CW) source. This parametric process facilitates the formation of mode-locked dissipative Kerr solitons (DKSs) through a double-balance mechanism: the dynamic equilibrium between nonlinear parametric gain and cavity loss, alongside the balance between self-phase modulation and cavity dispersion \cite{herr2014temporal,herr2016dissipative}. The presence of chromatic dispersion, however, introduces a frequency mismatch between the strictly equidistant comb lines and the non-equidistant cavity resonances. This mismatch, intertwined with nonlinear phase shifts, ultimately dictates the spectral bandwidth and power distribution of the microcomb \cite{godey2014stability}.

For pioneering applications such as chip-scale optical atomic clocks and precision frequency synthesizers, achieving an octave-spanning spectrum is imperative to enable direct $f$-$2f$ self-referencing \cite{hall2006nobel,hansch2006nobel,spencer2018optical,newman2019architecture,wu2025vernier}. Such a broad bandwidth is typically realized by engineering the resonator geometry to harness two key physical mechanisms: anomalous dispersion to shape the core soliton envelope, and the generation of dispersive waves (DWs) via soliton Cherenkov radiation to enhance the spectral wings \cite{brasch2016photonic}. Through precise geometric tailoring, octave-spanning microcombs can be directly generated on-chip \cite{li2017stably,pfeiffer2017octave}, bypassing the need for complex external spectral broadening links traditionally required in supercontinuum systems \cite{jost2015counting}.

Nevertheless, conventional dispersion engineering in uniform microring resonators relies on a fixed cross-section geometry. This constraint offers limited degrees of freedom and tightly couples higher-order dispersion coefficients to the secondary dispersion parameter, severely restricting the flexibility needed to optimize broadband dispersion profiles globally. While advanced strategies, such as in concentric rings \cite{kim2017dispersion}, dual-coupled cavities \cite{xue2015normal,yuan2023soliton,ji2026multicolor}, or photonic crystal resonators \cite{lucas2023tailoring}, can introduce localized dispersion perturbations via avoided mode-crossings (AMXs), the underlying mode hybridization inherently operates only within a narrow spectral window. Consequently, these localized approaches fall short of providing the global, ultra-broadband dispersion control required to shape and manipulate octave-spanning spectra. Although microwave-rate octave microcombs have recently been demonstrated using resonantly-coupled cavities to enhance the effective pump intensity \cite{zhu2026power}, achieving flexible and highly tailored global dispersion control remains an unresolved, fundamental challenge.

In this work, we report a comprehensive global dispersion engineering strategy based on heterostructured microresonators. By adiabatically concatenating multiple waveguide segments of distinct geometries, each with unique local dispersion, we synthesize tailored global dispersion profiles that transcend the limitations of uniform geometries. This strategy expands the dispersion tuning space by more than threefold compared to conventional designs. Capitalizing on this expanded design freedom, we demonstrate octave-spanning soliton microcombs from a single heterostructured microresonator, achieving repetition rates low enough for direct detection via commercial electronics.

\begin{figure*}[t!]
	\centering
	\includegraphics[width=0.9\textwidth]{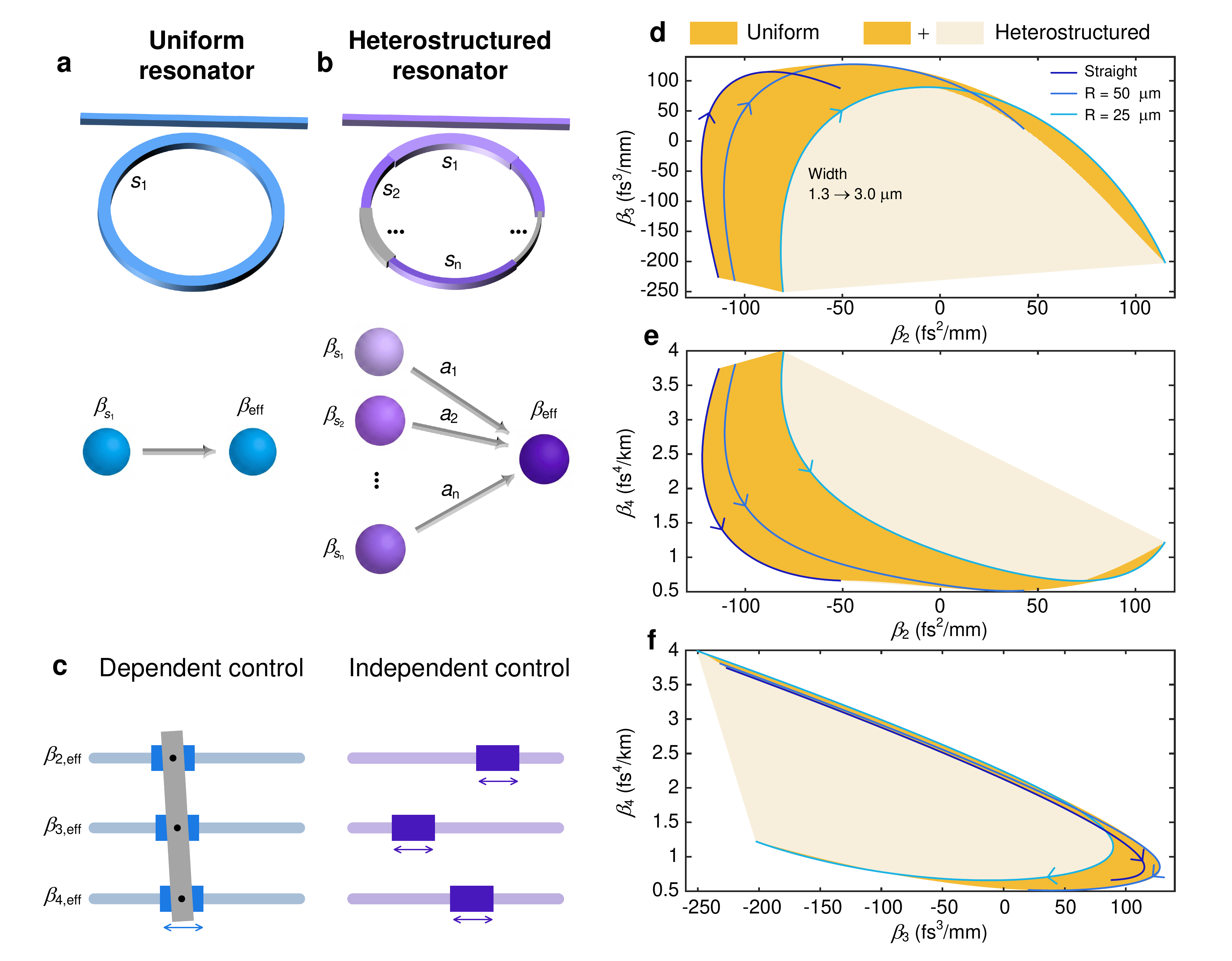}
	\caption{\textbf{Concept of tailoring resonator dispersion via a heterostructured architecture.} \textbf{a,} Schematic of a conventional uniform resonator, where the global effective dispersion is dictated by a single geometric structure. \textbf{b,} Schematic of a heterostructured resonator, synthesizing a customized $\beta_{\text{eff}}$ via a linearly weighted combination of multiple distinct substructures $(s_1,s_2,... s_n)$. \textbf{c,} Schematic of dispersion tuning constraints. Interlocked mechanical sliders illustrate the rigidly dependent parameter coupling in uniform resonators, contrasting with the independent control enabled by heterostructures. \textbf{d-f,} Comparison of accessible dispersion design spaces projected onto the $(\beta_2,\beta_3)$, $(\beta_2,\beta_4)$, and $(\beta_3,\beta_4)$ planes. Simulations are based on an 800-nm-thick $\text{Si}_3\text{N}_4$ core with a $\text{SiO}_2$ cladding. To establish a rigorous comparison, the heterostructured resonator is constructed from an arbitrary combination of substructures that share an identical geometric parameter space with the standard uniform ring resonator. Within this space, the waveguide width is varied from $1.3\,\upmu\text{m}$ to $3\,\upmu\text{m}$, and the bending radius ranges from $25\,\upmu\text{m}$ to infinity. Dark yellow regions represent the area attainable by uniform resonators, while light yellow regions illustrate the expanded accessible area achieved by heterostructured resonators. Solid curves trace the continuous dispersion trajectories of uniform waveguides at specific fixed bending radii.}
	\label{fig1}
\end{figure*}

\medskip
\noindent\textbf{Results}

\noindent\textbf{Dispersion engineering with a heterostructured resonator.} The conceptual framework of the heterostructured resonator, along with its comparison to conventional designs, is schematically illustrated in Fig. \hyperref[fig1]{1a,b}. This architecture is composed of multiple waveguide segments, each featuring distinct geometric parameters to leverage their unique local dispersion characteristics for custom global dispersion synthesis. Let $\beta_{m,s_i}$ denote the $m$-th order dispersion coefficient of the $i$-th segment $s_i$, the effective global dispersion is given by the weighted average of $\beta_{m,s_i}$, as follows: 
\begin{equation}
	\beta_{m,\text{eff}} = \sum_{i=1}^{n} a_i \beta_{m,s_i}\label{eq1}
\end{equation}
where $a_i = L_{s_i}/L_{\text{total}}$; $L_{s_i}$ is the segment length; $L_{\text{total}} = \sum_{i=1}^{n} L_{s_i}$ is the total roundtrip length.

\begin{figure*}[t!]
	\centering
	\includegraphics[width=0.65\textwidth]{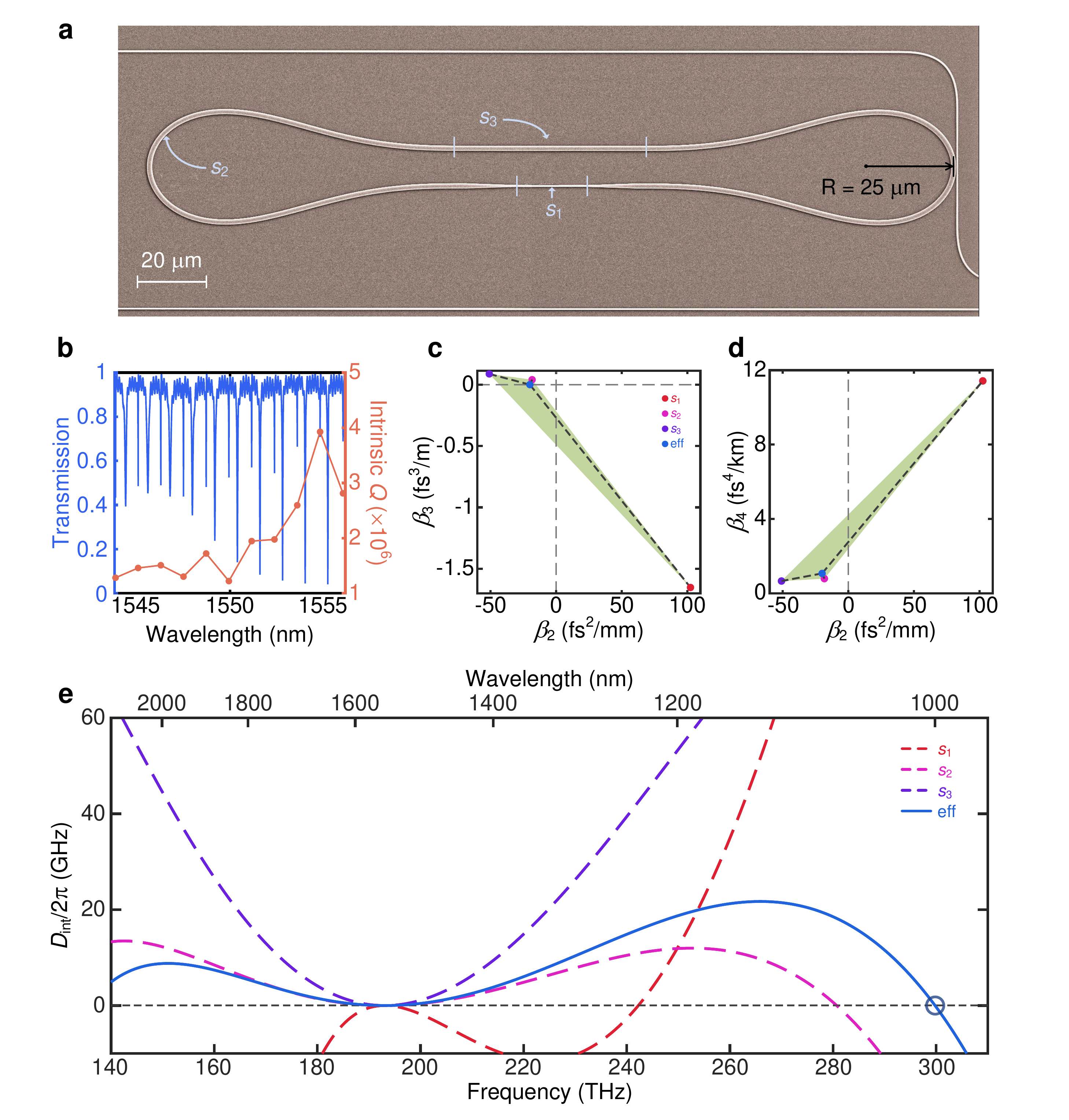}
	\caption{\textbf{Characterization of the heterostructured microresonator.} \textbf{a,} SEM image of the fabricated microresonator, which is composed of three distinct segments: segment $s_1$ is a 0.9-$\upmu$m-wide straight waveguide; segment $s_2$ is a 3.0-$\upmu$m-wide adiabatic bend featuring a quadratically varying curvature; and segment $s_3$ is a 3.0-$\upmu$m-wide straight waveguide. \textbf{b,} Measured transmission spectrum and the extracted intrinsic $Q$ factors. \textbf{c,d,} Designed accessible dispersion space projected onto the $(\beta_2,\beta_3)$ and $(\beta_2,\beta_4)$ planes. The shaded regions define the accessible dispersion area bounded by the specific structural coordinates. \textbf{e,} Synthesized global effective dispersion curve (solid line) alongside individual dispersion curves (dashed lines) corresponding to each substructure.}
	\label{fig2}
\end{figure*}

\begin{figure*}[t!]
	\centering
	\includegraphics[width=0.7\textwidth]{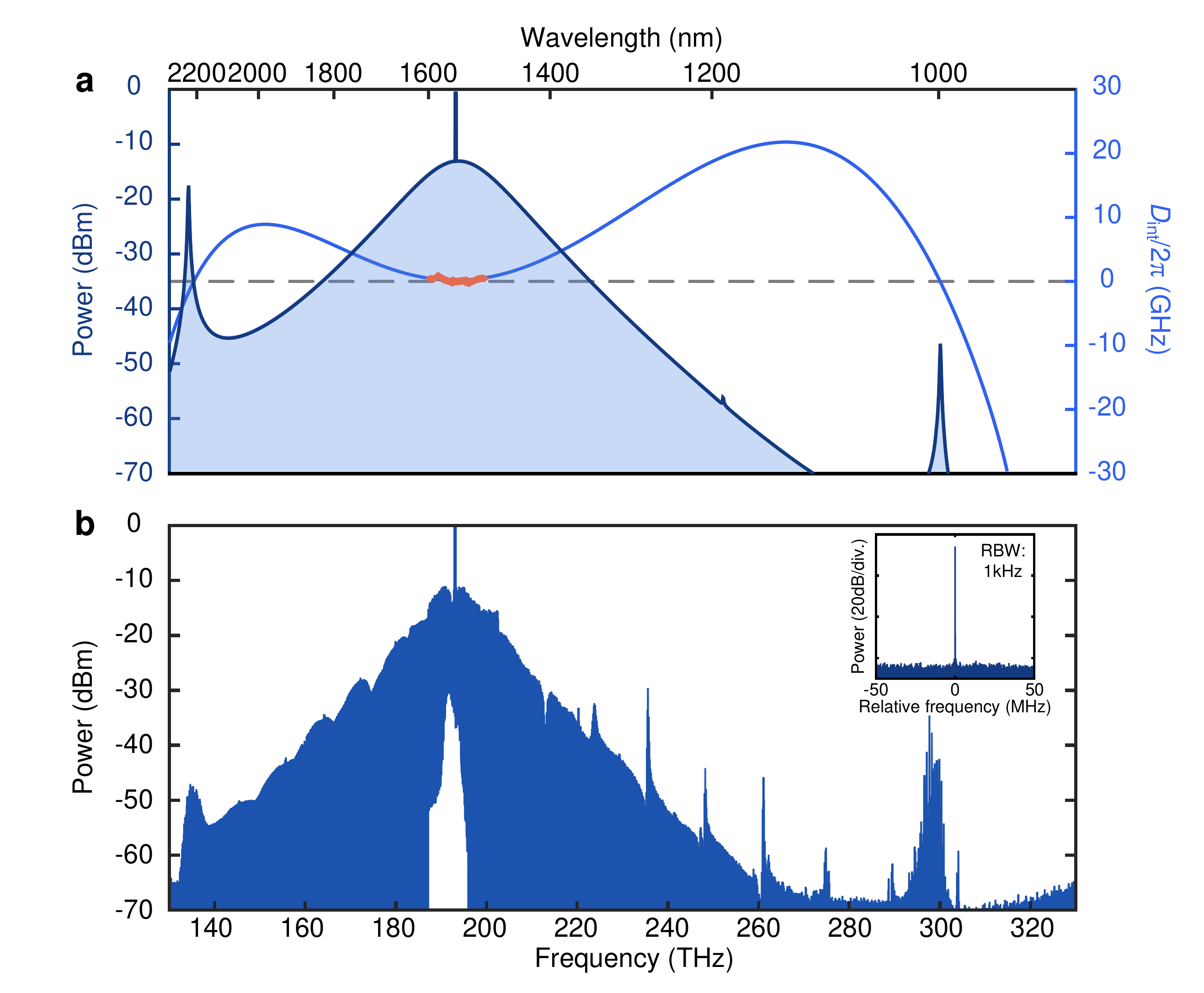}
	\caption{\textbf{Octave-spanning optical frequency comb generation via a heterostructured microresonator (MR1).} \textbf{a,} Designed integrated dispersion profile (light blue solid line) plotted alongside the measured dispersion data (orange dots), superimposed on the numerically simulated single-soliton spectrum (shaded area). \textbf{b,} Experimentally measured single-soliton spectrum. The inset displays the RF spectrum of the repetition rate beat note at 123.96 GHz, captured via heterodyne down-mixing using an electro-optic intensity modulator.}
	\label{fig3}
\end{figure*}

In conventional uniform microresonators, target values for the free spectral range (FSR) and second-order dispersion ($\beta_2$) are typically achieved by co-optimizing the bending radius and waveguide width once the waveguide height is fixed. However, during this geometric tuning process, higher-order dispersion coefficients shift inherently with $\beta_2$, precluding the capability for independent parameter control. In sharp contrast, heterostructured microresonators bypass this rigid coupling constraint by incorporating multiple distinct dispersive segments. By strategically adjusting the fractional lengths of these elements, higher-order dispersions can be engineered independently while anchoring the desired $\beta_2$, thereby unlocking a vastly expanded and highly customizable design parameter space.

To visually elucidate the disparity in dispersion-engineering capabilities between the two architectures, we calculated the dispersion profiles of waveguides across various geometric parameters using finite-difference eigenmode simulations based on an 800-nm-thick $\text{Si}_3\text{N}_4$ core with $\text{SiO}_2$ cladding (details regarding the simulation methods and parameters are provided in the Supplementary Section 1). As illustrated in Fig. \ref{fig1}, the simulated dispersion coefficients of different orders are projected onto the orthogonal coordinate axes, specifically mapping the $(\beta_2,\beta_3)$, $(\beta_2,\beta_4)$, and $(\beta_3,\beta_4)$ planes, to highlight the design contrast between uniform and heterostructured cavities. For a uniform ring resonator, sweeping the waveguide width at a fixed bending radius merely traces out a one-dimensional (1D) trajectory, delineating the rigid coupling between the mapped dispersion coefficients. Even as the bending radius varies, these trajectories only span a bounded two-dimensional (2D) region, which defines the strict boundary of the accessible dispersion space for uniform rings. Conversely, for a heterostructured microresonator, the global effective dispersion is governed by a linearly weighted combination of its constituent substructures. Mathematically, this implies that within these 2D parameter planes, any arbitrary point situated on the line connecting two sub-structural coordinates can be synthesized. Consequently, the achievable parameter space fundamentally expands to form a complete convex hull, effectively filling and extending far beyond the outermost boundaries of uniform resonators.

\begin{figure*}[t!]
	\centering
	\includegraphics[width=0.66\textwidth]{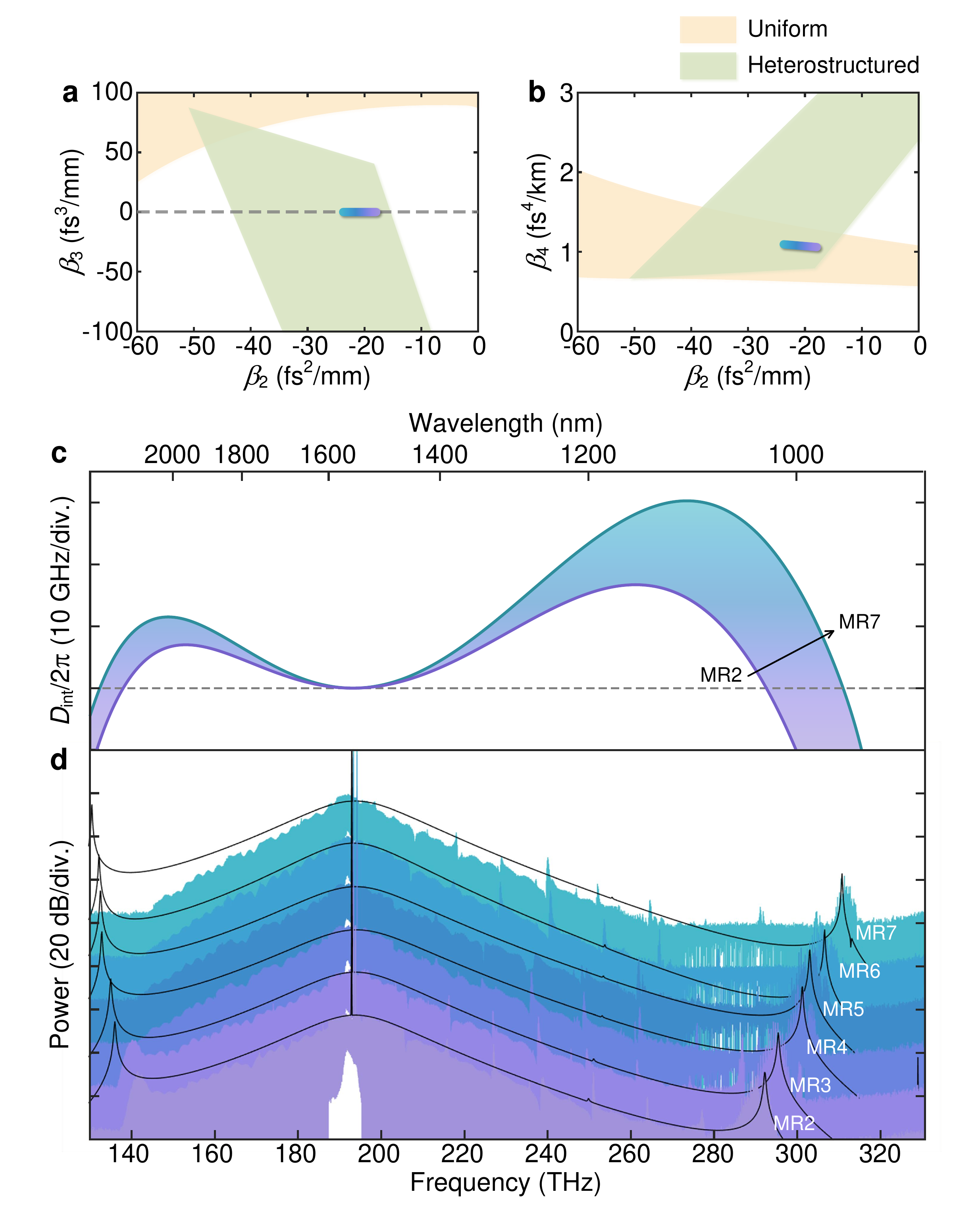}
	\caption{\textbf{Spectral tuning of dispersive waves enabled by the heterostructured architecture.} \textbf{a,b,} Calculated accessible dispersion regions for uniform (orange shaded areas) and heterostructured (green shaded areas) microresonators. The solid line, color-coded with a cyan-to-purple linear gradient, traces the trajectory of dispersion parameters during dispersive wave (DW) tuning. \textbf{c,} Numerically simulated integrated dispersion curves for microresonators MR2 to MR7. \textbf{d,} Corresponding experimental optical spectra of the generated single-soliton states in MR2 to MR7, with the theoretical simulation results superimposed as matching solid lines for comparison.}
	\label{fig4}
\end{figure*}

To quantify this enhancement, we evaluated the accessible dispersion space ratio between the two architectures. Notably, the uniform microring space accounts for only 34\%, 44\%, and 15\% of the total area accessible by the heterostructured counterparts in the $(\beta_2,\beta_3)$, $(\beta_2,\beta_4)$, and $(\beta_3,\beta_4)$ planes, respectively. Beyond significantly expanding the design area, the heterostructured architecture exhibits geometric degeneracy, meaning a target $(\beta_2,\beta_3,\beta_4)$ dispersion combination can be synthesized via multiple distinct substructure configurations. This decoupling capability enables independent, multilevel optimization of other critical factors governing microcomb dynamics, such as the FSR and the coupling coefficient, while anchoring the target dispersion profile. Consequently, this architecture establishes a versatile paradigm for precision dispersion engineering tailored to diverse application scenarios.

\medskip
\noindent\textbf{Octave-spanning microcomb generation.} We designed and fabricated a batch of heterostructured microresonators (designated as MR1$\sim$MR7) tailored for octave-spanning microcomb generation, with repetition rates ranging from approximately 118 GHz to 124 GHz. The photonic chips were fabricated using a standard commercial foundry process (LIGENTEC), featuring an 800-nm-thick $\text{Si}_{3}\text{N}_4$ waveguide core embedded in a $\text{SiO}_{2}$ cladding. A scanning electron microscope (SEM) image of a representative resonator is shown in Fig. \hyperref[fig2]{2a}. Structurally, the microresonator incorporates three distinct waveguide geometries as constituent components: a 0.9-$\upmu\text{m}$-wide straight waveguide serves as segment $s_1$ to suppress AMXs between the fundamental and higher-order modes; a 3.0-$\upmu\text{m}$-wide adiabatic bend functions as segment $s_2$ to ensure broadband coupling with the bus waveguide. This adiabatic section tapers from a 25-$\upmu\text{m}$ bending radius to a straight waveguide section, utilizing a curvature that varies quadratically with the arc length. This quadratic profile induces weaker mode coupling compared to conventional Euler bends \cite{chen2012general} (detailed characteristics of the adiabatic bend are provided in the Supplementary Section 4). Finally, a 3.0-$\upmu\text{m}$-wide straight waveguide serves as segment $s_3$, acting as the tuning element to globally balance the cavity dispersion.

Systematically adjusting the fractional weights of $s_1$, $s_2$, and $s_3$ enables the synthesis of diverse effective dispersion profiles. The accessible dispersion space forms a triangular region bounded by these three structural coordinates (Fig. \hyperref[fig2]{2c,d}). The effective second-, third-, and fourth-order dispersion coefficients are evaluated as follows:
\begin{equation}
	\begin{bmatrix}
		\beta_{2,\text{eff}} \\
		\beta_{3,\text{eff}} \\
		\beta_{4,\text{eff}}
	\end{bmatrix}
	=
	\begin{bmatrix}
		\beta_{2,s_1} & \beta_{2,s_2} & \beta_{2,s_3} \\
		\beta_{3,s_1} & \beta_{3,s_2} & \beta_{3,s_3} \\
		\beta_{4,s_1} & \beta_{4,s_2} & \beta_{4,s_3}
	\end{bmatrix}
	\begin{bmatrix}
		a_1 \\
		a_2 \\
		a_3
	\end{bmatrix}\label{eq2}
\end{equation}

To solve for the optical weight parameters, we applied a multi-objective constraint framework: first, we set $\beta_{3,\text{eff}}$   = 0 and assign opposite signs to $\beta_{2,\text{eff}}$  and $\beta_{4,\text{eff}}$. Under zero third-order dispersion, the interplay between second- and fourth-order dispersions naturally yields two DWs at opposite spectral ends, satisfying the phase-matching condition $D_{\text{int}}(\mu) = 0$. Here, the integrated dispersion is defined as $D_{\text{int}}(\mu) = \omega_\mu  -\omega_0 - D_1\mu$, where $\mu$ is the relative mode number and $D_1/2\pi$ denotes the FSR (the analytical formulation relating the integrated dispersion of the heterostructured resonator to its constituent segments is detailed in the Supplementary Section 2). Second, we constrained a target DW to emerge precisely at a specific self-referencing frequency (e.g., 300 THz) to locally boost the comb line power. Third, the standard normalization condition $\sum a_i = 1$ was enforced. These combined constraints uniquely determine the structural weight parameters, with an exemplary engineered dispersion profile illustrated in Fig. \hyperref[fig2]{2e}.

The transmission spectrum of the microresonator was characterized by scanning a tunable CW laser and monitoring the transmitted optical power \cite{li2012sideband} (see methods for details). By measuring and fitting the resonance frequencies along with their full width at half maximum (FWHM) linewidths, we extracted the cavity quality ($Q$) factors and the integrated dispersion profile. For MR1, the measured transmission spectrum across the 1545-1555 nm band and the intrinsic $Q$ factor are plotted in Fig. \hyperref[fig2]{2b}. The corresponding measured $D_\text{int}$ and the theoretical dispersion curve are displayed in Fig. \hyperref[fig3]{3a}. The measured data are in good agreement with the theoretical curves.

The experimental setup for microcomb generation is detailed in the Methods section. An auxiliary laser was introduced as a cooling light to mitigate the adverse effects of thermal instabilities within the microresonator \cite{zhou2019soliton}. In the experiment, the on-chip pump power and the auxiliary laser power were 770 mW and 740 mW, respectively. The experimental single-soliton spectrum for MR1 is shown in Fig. \hyperref[fig3]{3b}. The corresponding simulation results are displayed in Fig. \hyperref[fig3]{3a}, with the numerical approach detailed in the Methods section. The spectral envelope matches the profile of an anomalous dispersion soliton superimposed with DWs. Within the spectrum, localized power discontinuities are observed, caused by the residual AMXs between the fundamental and higher-order modes. The generated DWs appear near 130 THz and 300 THz. Additionally, the microcomb repetition rate was down-converted using an electro-optic intensity modulator and detected via a commercial photodetector. The results are displayed in the inset of Fig. \hyperref[fig3]{3b}. The sharp spectral line confirms a low-noise mode-locked soliton state.

\medskip
\noindent\textbf{Dispersive wave tuning.} Under the constraint of a zero third-order dispersion ($\beta_{3,\text{eff}} = 0$), regulating $\beta_{2,\text{eff}}$ enables precise tuning of the DW positions, which is realized by systematically varying the weight relationships among $s_1$, $s_2$, and $s_3$. In Fig. \hyperref[fig4]{4a,b}, the deterministic tuning of $\beta_{2,\text{eff}}$ is illustrated by solid lines color-coded with a linear gradient. Remarkably, these optimized dispersion parameters lie beyond the reach of conventional uniform architectures. Fig. \hyperref[fig4]{4c} presents the integrated dispersion curves for DW tuning, demonstrating a flexible spectral tuning range spanning from 290 THz to 310 THz. To validate this framework, we fabricated these microresonators (designated as MR2 to MR7) and performed systematic microcomb generation experiments. The resulting single-soliton spectra are displayed in Fig. \hyperref[fig4]{4d}, where a larger anomalous dispersion systematically pushes the DWs further away from the central pump frequency.

The experimental single-soliton spectral envelopes exhibit excellent agreement with the theoretical simulations on the short-wavelength side. On the long wavelength side, however, the measured comb power undergoes a noticeable roll-off, and certain spectra lack the expected long-wavelength DW peak. This discrepancy primarily stems from the elevated propagation losses of the 0.9-$\upmu$m-wide intracavity straight waveguide segment ($s_1$) within the long-wavelength regime. Nevertheless, these experimental findings firmly demonstrate that our heterostructured architecture is capable of generating octave-spanning microcombs while simultaneously enabling agile and highly flexible tuning of the DW positions. The close match between the target dispersion designs and the measured optical spectra validates the robustness and efficacy of this approach for advanced, multi-degree-of-freedom dispersion engineering.

\begin{figure}[h]
	\centering
	\includegraphics[width=0.5\textwidth]{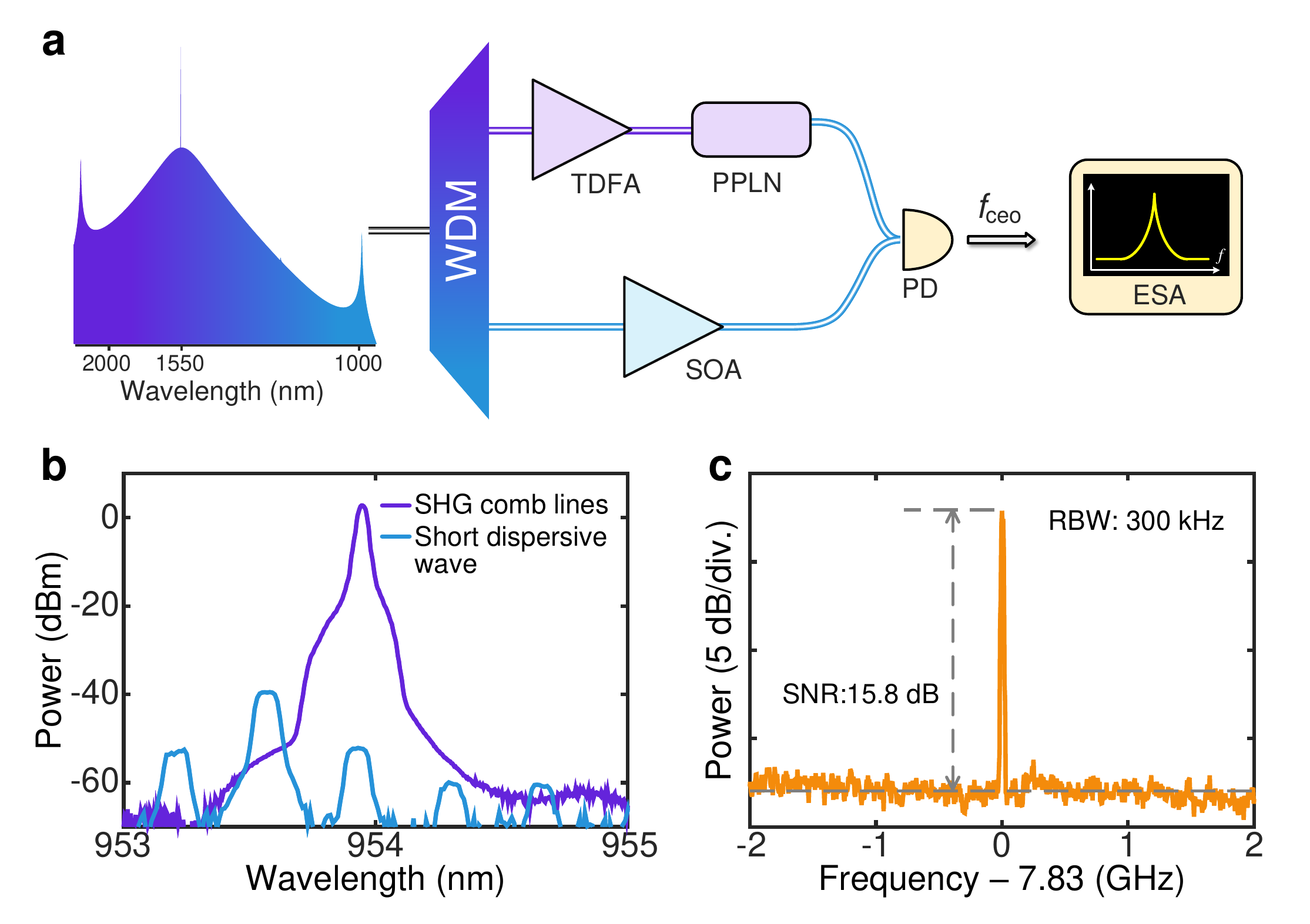}
	\caption{\textbf{Detection of the carrier-envelope offset frequency.} \textbf{a,} Schematic of the experimental setup for $f$-$2f$ self-referencing interferometry. WDM, wavelength-division multiplexer; TDFA, thulium-doped fiber amplifier; PPLN, periodically poled lithium niobate; SOA, semiconductor optical amplifier; PD, photodetector; ESA, electrical spectrum analyzer. \textbf{b,} Overlapping optical spectra of the generated second-harmonic signal and the short-wavelength dispersive wave comb lines. \textbf{c,} Radio-frequency spectrum of the detected $f_\text{ceo}$ beat note, exhibiting a signal-to-noise ratio of 15.8 dB recorded at a resolution bandwidth of 300 kHz.}
	\label{fig5}
\end{figure}

\medskip
\noindent\textbf{Carrier-envelope offset frequency characterization.} To characterize the coherence of the generated DWs, we constructed an $f$-$2f$ self-referencing interferometry setup to detect the carrier-envelope offset frequency ($f_\text{ceo}$), as schematically illustrated in Fig. \hyperref[fig5]{5a}. A wavelength division multiplexer (WDM) splits the broadband microcomb into long- and short-wavelength arms, corresponding to the $f$ and $2f$ spectral regions, respectively. Within the long-wavelength arm, specific comb lines are selected by a tunable bandpass filter, amplified via cascaded thulium-doped fiber amplifiers (TDFAs), and subsequently focused into a periodically poled lithium niobate (PPLN) crystal to undergo second-harmonic generation (SHG). Concurrently, the short-wavelength comb lines are independently amplified by a semiconductor optical amplifier (SOA). These amplified lines are then combined with the generated SHG signal and directed into a fast PD to produce the $f_\text{ceo}$ beat note.

The overlapping optical spectra of the SHG signal (derived from the long-wavelength comb lines) and the short-wavelength DW comb lines are shown in Fig. \hyperref[fig5]{5b}. The $f_\text{ceo}$ beat note measured at a resolution bandwidth (RBW) of 300 kHz is displayed in Fig. \hyperref[fig5]{5c}, exhibiting a signal-to-noise ratio (SNR) of approximately 15.8 dB. In the future, the SNR may be further elevated by optimizing the on-chip bus-ring coupling architecture to enhance the comb extraction efficiency \cite{moille2019broadband}.

\medskip
\noindent\textbf{Discussion}

\noindent In summary, we have experimentally demonstrated a global dispersion synthesis paradigm utilizing heterostructured microresonators, and successfully realized single-chip, octave-spanning DKS microcombs. The custom-engineered dispersion profiles facilitate detectable repetition rates ranging from 118 to 124 GHz, featuring tailorable dispersive waves in the 290 to 310 THz range. Consequently, the direct detection of the $f_\text{ceo}$ frequency via $f$-$2f$ self-referencing was achieved with a signal-to-noise ratio of 15.8 dB.

Notwithstanding the outstanding design flexibility afforded by this architecture, several avenues remain open for further optimization. Currently, an auxiliary laser is employed to counteract the thermal instabilities during soliton activation. This overhead could be bypassed through advancements in ultra-low-loss nanofabrication to drastically suppress absorption-induced heating \cite{ji2025deterministic}. Furthermore, translating the heterostructured concept into coupled multi-cavity platforms holds immense potential to significantly enhance pump-to-comb conversion efficiency \cite{zhu2026power,helgason2023surpassing}. Such power-efficient operation would pave the way for heterogeneous integration with pump laser diodes, ultimately unleashing turnkey, self-injection-locked, octave-spanning microcombs \cite{briles2021hybrid,shen2020integrated,wildi2024phase,ulanov2024synthetic}. On the design front, the exploration of microresonator geometries can be revolutionized by integrating advanced inverse-design techniques and machine-learning algorithms \cite{lucas2023tailoring,molesky2018inverse}. These automated computational frameworks will systematically navigate the highly degenerate, multidimensional parameter space to uncover globally optimized structural configurations. Beyond octave-spanning microcombs, the proposed approach can be readily extended to synthesize near-zero dispersion over a broad bandwidth for ultra-flat comb generation (see Supplementary Section 3), which holds great promise for high-capacity optical communications.

\medskip
\noindent\textbf{Methods}

\begin{footnotesize}

\noindent\textbf{Frequency-domain microcomb simulation.} Numerical simulations of the microcombs in the frequency domain were performed based on the Lugiato-Lefever equation (LLE) under the mean-field approximation \cite{lugiato1987spatial,coen2012modeling}:
\begin{equation}
	\frac{\partial \tilde{A}_{\mu}}{\partial t} = -\left[ \frac{\kappa}{2} + i\delta\omega - iD_{\text{int}}(\mu) \right] \tilde{A}_{\mu} + ig_0 \mathcal{F}[|A|^2 A]_{\mu} + \sqrt{\frac{\kappa_{\text{ex}} P_{\text{in}}}{\hbar\omega_0}} \delta_{\mu=0}\label{eq3}
\end{equation}
where $\mu$ is the azimuthal mode number relative to the pumped mode, $|\tilde{A}_{\mu}|^2$ represents the number of photons in mode $\mu$ as a function of time $t$, $P_{\text{in}}$ and $\delta\omega = \omega_0 - \omega_{\text{pump}}$ are the pump power and detuning, respectively. $\kappa$ is the FWHM linewidth, and $\kappa_{\text{ex}}$ is the external coupling rate. The parameter $g_0$ is the per-photon Kerr shift, and $\mathcal{F}[\cdot]_{\mu}$ denotes the Fourier series operator.

The resonator parameters in the simulation were derived from the measured data, where $\kappa \approx 2\pi \times 1.51 \, \text{GHz}$ and $\kappa_{\text{ex}} \approx 2\pi \times 1.35 \, \text{GHz}$. The nonlinear refractive index $n_2$ for $\text{Si}_3\text{N}_4$ was set to about $2.4 \times 10^{-19} \, \text{m}^2/\text{W}$, yielding a corresponding $g_0 \approx 2\pi \times 0.35 \, \text{Hz}$. The detuning was treated as an adjustable parameter to match the experimental results in Fig. \ref{fig3} and Fig. \ref{fig4}.

\medskip
\noindent\textbf{Experimental setup for transmission spectrum measurement.} A tunable external cavity diode laser (ECDL, Keysight 81606A) operating in a continuous frequency sweep mode (spanning 1500\textendash{}1600 nm) was employed to acquire the linear transmission spectra. The laser output was split into two parallel paths using an optical coupler: one portion was launched into the microresonator device under test and subsequently detected by a PD; the other portion was routed to a pre-calibrated, unbalanced Mach-Zehnder interferometer (MZI) and detected by a second PD. The unbalanced MZI trace was utilized for the real-time calibration of the ECDL's frequency sweep nonlinearity, thereby enabling the highly precise extraction of the resonance frequencies and their FWHM linewidths. The MZI features a FSR of approximately 10 MHz, providing sufficient resolution to accurately calibrate resonance peaks with a FWHM greater than 100 MHz.

\medskip
\noindent\textbf{Experimental setup for microcomb generation.} Two independent ECDLs (Toptica DLC pro and Keysight 81606A) were deployed as the pump and auxiliary cooling lasers, respectively, during the microcomb generation process. Both lasers were boosted by high-power erbium-doped fiber amplifiers (EDFAs), combined via a polarization beam splitter, and coupled into the microresonator device. The pump laser was aligned to excite the transverse electric (TE) mode, while the auxiliary laser was coupled into the transverse magnetic (TM) mode. Experimentally, the pump and auxiliary wavelengths were parked on the red-detuned side of a TE resonance and the blue-detuned side of a TM resonance, respectively. By sweeping the voltage applied to the integrated microheater, the effective pump-to-resonance detuning was dynamically tuned via the thermo-optic effect, successfully triggering the generation of DKSs. Backward tuning of the pump was utilized to reduce the soliton number until a single soliton state was achieved \cite{guo2017universal}. The broadband microcomb output was separated into long and short wavelength bands using a WDM and recorded by two distinct optical spectrum analyzers (OSAs, Yokogawa AQ6370D and AQ6375E). A small tap fraction ($\sim 1\%$) of the total output was diverted to a fast oscilloscope to monitor the characteristic soliton steps in real-time, and simultaneously to an ESA to verify the coherence of the generated combs.

\end{footnotesize}
%

\noindent\textbf{Author contributions} 

\begin{footnotesize}
\noindent The idea and the experimental setup were conceived and designed by X.L. and X.X. Measurements and data analysis were performed by X.L., with assistance from Y.L. and Z.W. Electron microscopy imaging was conducted by W.Z. Numerical simulations and analysis were carried out by X.L., with the help of Q.X. All authors participated in preparing the manuscript.
\end{footnotesize}

\medskip

\noindent\textbf{Competing interests}

\begin{footnotesize}
\noindent The authors declare no competing interests.
\end{footnotesize}

\bibliography{refbase.bib}

\medskip

\noindent\textbf{Data availability}

\begin{footnotesize}
\noindent The data that support the plot within this paper and other findings of this study are available upon publication. 
\end{footnotesize}

\medskip

\noindent\textbf{Code availability}

\begin{footnotesize}
\noindent The codes that support the findings of this study are available upon publication. 
\end{footnotesize}

%
%

\end{document}